\documentclass[cameraready]{Interspeech}

\title{FoleyGenEx: Unified Video-to-Audio Generation with Multi-Modal Control, Temporal Alignment, and Semantic Precision}

\author[affiliation={1}, orcid=0009-0006-5786-7109, equalcontribution]{Shiyao}{Wang}
\author[affiliation={2}, orcid=0009-0002-3668-2392, equalcontribution]{Xijuan}{Zeng}
\author[affiliation={1}, orcid=0009-0003-8057-4644, equalcontribution]{Hui}{Wang}
\author[affiliation={1},orcid=0000-0001-5068-025X]{Shiwan}{Zhao}
\author[affiliation={2}]{Feng}{Deng}
\author[affiliation={2}]{Chen}{Zhang}
\author[affiliation={1}, orcid=0009-0000-2748-3020, correspondingauthor]{Yong}{Qin}

\address{
  $^1$ Academy for Advanced Interdisciplinary Studies, Nankai University, Tianjin, China \\
  $^2$ Kling Team, Kuaishou Technology, Beijing, China 
  \thanks{This work was done during Shiyao Wang's internship at Kuaishou Technology.}
}

\email{wangshiyao@mail.nankai.edu.cn, qinyong@nankai.edu.cn}

\usepackage{comment}

\begin{document}

\maketitle

\begin{abstract}
    We present FoleyGenEx, a unified video-to-audio (VTA) framework integrating multi-modal control, frame-level temporal alignment, and fine-grained semantics, enabling synchronized, versatile audio synthesis for diverse tasks. Existing VTA methods either have multi-modal control but weak temporal alignment or strong alignment but lack reference audio conditioning and semantic precision. FoleyGenEx fills this gap via three core innovations: a conditional injection mechanism for audio-controlled VTA and Foley extension, a multi-modal dynamic masking strategy preserving training synchronization, and an adverb-based data augmentation algorithm leveraging signal processing and large language models to enhance textual supervision with nuanced semantics. Experiments on AudioCaps, VGGSound, and Greatest Hits demonstrate its competitive controllable VTA performance against existing methods. Demo samples are available at \url{https://foleygenex.github.io/FoleyGenEx}.
\end{abstract}

\section{Introduction}
Text-to-video (T2V) generation has made remarkable progress with models such as OpenAI’s Sora \cite{sora} and others \cite{cogvideox, hunyuanvideo, kling, wan}, producing visually compelling content. However, most T2V outputs remain silent, severely limiting user immersion and realism. Adding synchronized audio would dramatically enhance the viewing experience, yet manual dubbing and alignment are prohibitively time-consuming. This challenge has driven growing interest in generative models for video-to-audio (VTA) synthesis, which aim to automatically produce audio tracks conditioned on video content.

Despite recent advances, existing VTA methods\cite{V-AURA,VATT,Frieren,foleycrafter,V2A-Mapper,klingfoley,multifoley,MMAudio} face persistent challenges in three key areas: temporal synchronization, multi-modal control, and fine-grained semantic precision. MultiFoley \cite{multifoley} integrates video, text, and reference audio by upsampling video semantic features and performing channel-wise concatenation with the reference audio in the latent space to enable style transfer and Foley extension (FE). However, this simplistic upsampling strategy often results in degraded temporal synchronization.In contrast, MMAudio \cite{MMAudio} achieves superior frame-level synchronization by leveraging a Multi-modal Diffusion Transformer (MMDiT) \cite{sd3} architecture. This framework effectively disentangles different modalities and facilitates complex cross-modal interactions through joint-attention mechanisms, further enhanced by the Synchformer \cite{synchformer} for precise alignment. Despite its synchronization strengths, MMAudio lacks a dedicated reference audio conditioning branch. Consequently, it struggles with tasks like audio-controlled VTA (AC-VTA) and FE, where the objective is to generate audio that precisely matches the timbre, prosody, and acoustic events of a reference track. Moreover, both approaches fail to provide fine-grained semantic control when textual descriptions specify subtle variations in manner or intensity—for example, distinguishing between ``fast knocking" and ``slow knocking," or ``loud knocking" and ``soft knocking"—largely because standard datasets \cite{vggsound,audiocaps,wavcaps} contain few adverbial cues, offering limited supervision for such nuanced semantics.

\begin{figure}[t]
\begin{center} 
\setlength{\abovecaptionskip}{0.0cm} 
\centerline{\includegraphics[width=8cm]{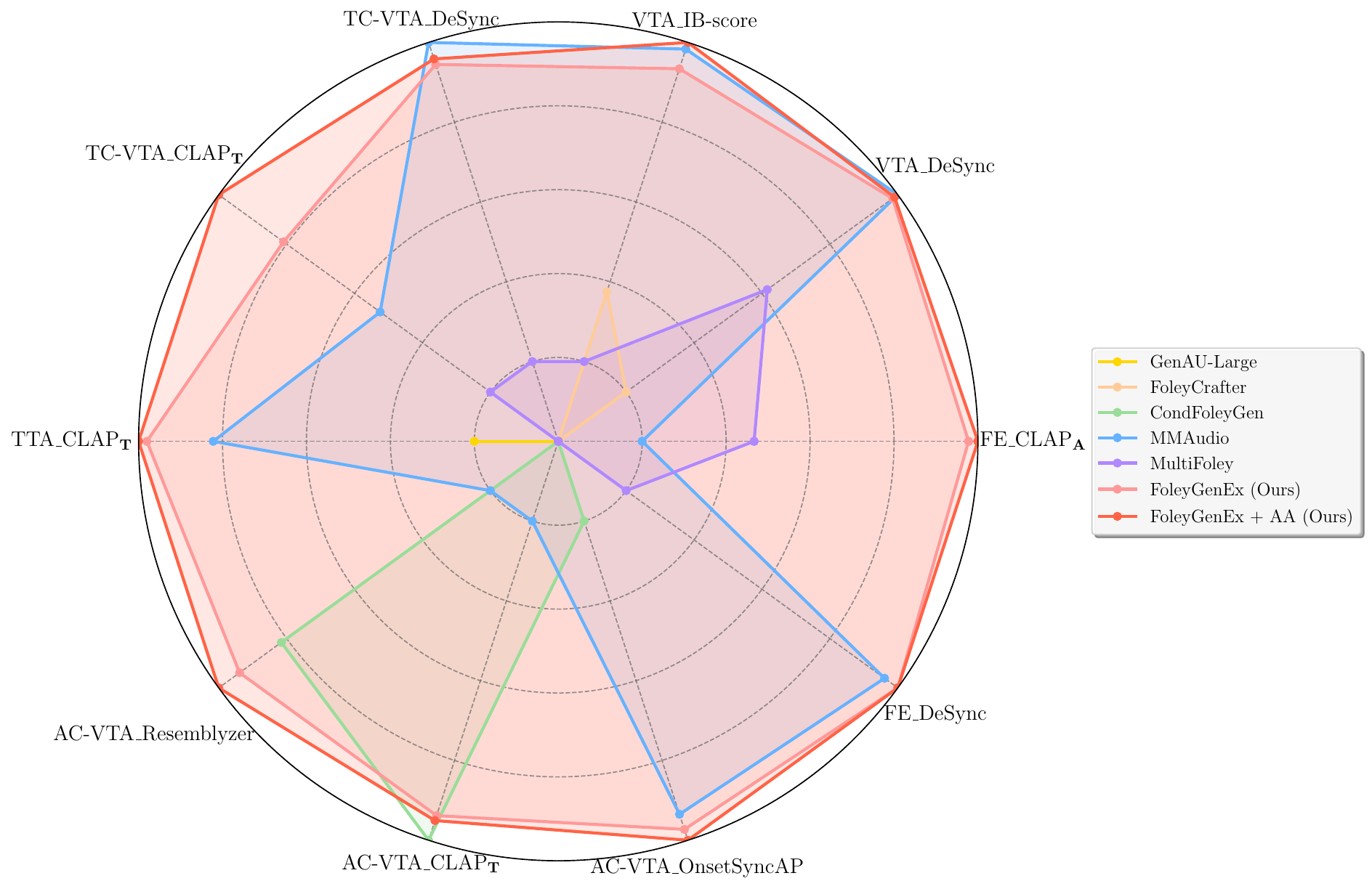}}
\captionof{figure}{FoleyGenEx supports a range of multi-modal controlled audio generation tasks, including Text-to-Audio (TTA), basic Video-to-Audio (VTA), Text-Controlled VTA (TC-VTA), Audio-Controlled VTA (AC-VTA), and Foley extension (FE). It unifies these tasks while achieving strong synchronization, versatile control, and expressive audio generation.}
\label{f5:radar}
\vspace{-39pt}
\end{center}
\end{figure}

These limitations create a clear trade-off: MultiFoley supports diverse control inputs but sacrifices temporal precision, whereas MMAudio delivers strong synchronization but lacks versatility in conditioning and semantic expressivity. As illustrated in Figure \ref{f5:radar}, neither approach achieves all three objectives simultaneously, leaving a gap for methods that can unify temporal alignment, multi-modal control, and semantic precision within a single framework.

\begin{figure*}[t]
\setlength{\abovecaptionskip}{0.2cm} 
\centerline{\includegraphics[width=15cm]{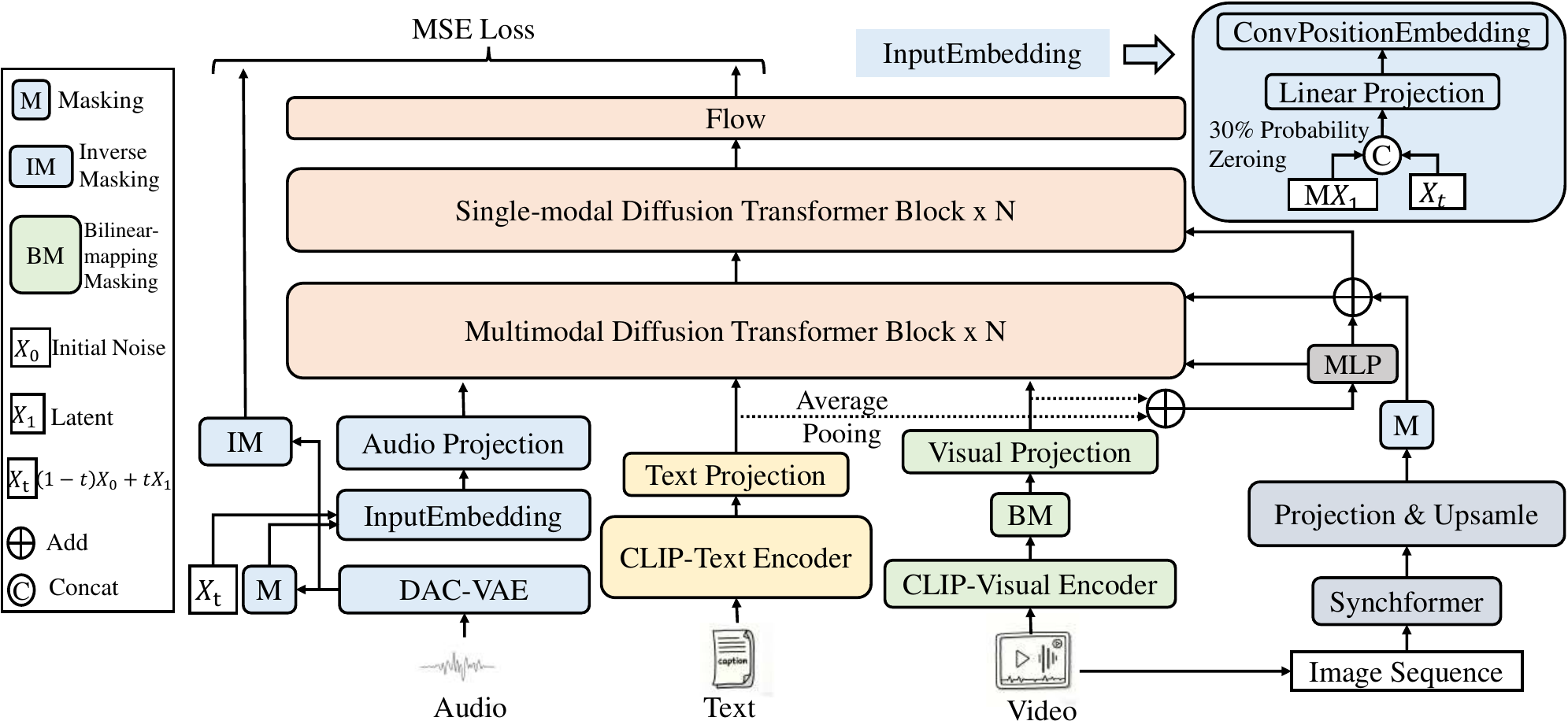}}
\caption{FoleyGenEx training framework.}
\label{f2:training framework}
\vspace{-20pt}
\end{figure*}
To close this gap, we introduce FoleyGenEx, a novel VTA framework built upon the MMDiT architecture. FoleyGenEx preserves the strong synchronization capabilities of MMAudio while extending its functionality with a conditional injection mechanism that enables reference audio conditioning for AC-VTA and FE. A multi-modal dynamic masking strategy further ensures alignment between training and inference, mitigating the synchronization degradation observed in MultiFoley. Beyond synchronization and control, FoleyGenEx incorporates an adverb-based data augmentation algorithm that leverages both signal processing techniques and large language models (LLMs) to enrich training data with adverbial cues, enabling fine-grained semantic control over the generated audio.

Extensive experiments on AudioCaps \cite{audiocaps}, VGGSound \cite{vggsound}, and Greatest Hits \cite{GreatestHits} demonstrate that FoleyGenEx not only inherits the strong synchronization performance of MMAudio but also achieves the versatile multi-modal control of MultiFoley, while uniquely providing fine-grained semantic precision through adverb-based augmentation. As shown in Figure \ref{f5:radar}, FoleyGenEx and its adverb-augmented variant achieve competitive performance compared to existing methods across synchronization, control, and semantic metrics, and excel in several key aspects, representing significant progress for controllable VTA generation.

\section{Related work}
\noindent\textbf{Flow matching} Flow matching \cite{flowmatching} models the continuous transformation of data distributions using an ordinary differential equation (ODE):
\begin{align}
    \frac{dX_{\tau}}{d\tau} = v(\tau, X_{\tau}),
\end{align}
where \(X_{\tau}\) represents the state at continuous time \( \tau \) (ranging from 0 to 1), and \(v(\tau, X_{\tau})\) is a learned flow field guiding \(X_{\tau}\) from an initial distribution \(p_0(X)\) (e.g., Gaussian noise) to a target distribution \(p_1(X)\) (the real data distribution). To solve this ODE numerically, Euler's method is used for discretization, resulting in the iterative update formula:
\begin{align}
    X_t = t \cdot X_1 + (1-t) \cdot X_0,
\end{align}
where \(t\) is a discrete time step (a scalar controlling interpolation), \(X_0\) is a sample from the initial noise distribution \(p_0(X)\), and \(X_1\) is a sample from the target data distribution \(p_1(X)\).

\noindent\textbf{Multi-modal diffusion transformer} MMDiT \cite{sd3} builds upon the Diffusion Transformer (DiT) architecture, emphasizing collaborative modeling and information interaction across multiple modalities. This is crucial for cross-modal generative tasks like VTA, which require coordinated multi-source conditioning. MMDiT uses separate weight parameters to construct independent feature streams for each modality. Within attention layers, it concatenates multi-modal sequences to enable bidirectional cross-modal information flow, preserving modality-specific characteristics while ensuring precise alignment. To further enhance cross-modal collaboration, the adaLN mechanism \cite{adaln} is integrated to inject conditions.

\noindent\textbf{Multi-modal controlled audio generation} Text-to-audio (TTA) \cite{audioldm2,tango2,makeanaudio2,GenAU-Large} offers control over background and environmental sounds but often struggles with video semantics or accurate synchronization. Conversely, VTA excels at generating audio aligned with video content and events, essential for applications like silent film dubbing and damaged audio track restoration. Recent efforts focus on multi-modal controlled audio generation to increase VTA's flexibility. For example, CondFoleyGen \cite{CondFoleyGen} utilizes a target video, a conditional reference video, and a conditional reference audio track to synthesize audio that is temporally synchronized with the target video while inheriting the specific acoustic style and timbre from the reference audio.
Sketch2Sound \cite{sketch2sound} uses loudness, spectral centroid, and pitch probabilities from reference audio to control the generated audio's prosody. MultiFoley integrates video, text, and audio by aligning upsampled video features with reference audio latents via channel-wise concatenation to achieve multi-modal control for style transfer and Foley extension.

\begin{figure*}[t]
\setlength{\abovecaptionskip}{0.1cm} 
\centerline{\includegraphics[width=15cm]{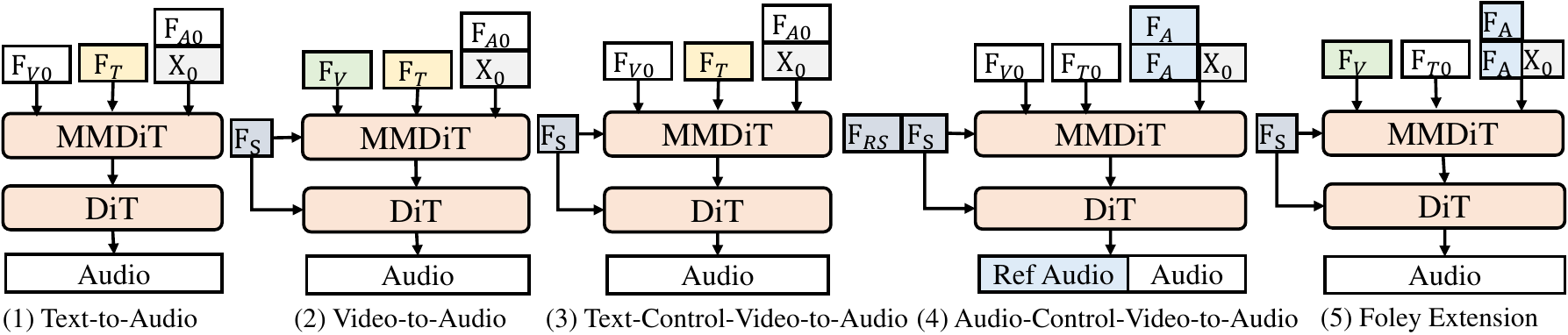}}
\caption{Multi-modal controlled audio generation tasks. F\textsubscript{T}: text semantic features; F\textsubscript{V}: video semantic features; F\textsubscript{S}: video synchronization features; F\textsubscript{RS}: reference video synchronization features; F\textsubscript{A}: reference audio latents. F\textsubscript{V0}, F\textsubscript{A0}, and F\textsubscript{T0} represent all-zero features for video, audio, and text, respectively. `Ref' denotes `reference'.}
\label{f3:inference-setup}
\vspace{-18pt}
\end{figure*}
\section{FoleyGenEx}
\label{sec:method}
\subsection{Overview}
\label{sec:overview}
The FoleyGenEx training framework, as depicted in Figure \ref{f2:training framework}, leverages an MMDiT backbone for multi-modal feature fusion and a single-modal DiT for high-fidelity audio modeling via iterative flow-matching. The core architectural innovation lies in our conditional injection mechanism within the audio stream, synergized with a novel multimodal dynamic masking strategy that ensures robust cross-modal alignment.

\noindent\textbf{Audio modality}
Audio latents are extracted using a pre-trained DAC-VAE \cite{dacvae}. To facilitate conditional injection and ensure zero-shot generalization, we employ a stochastic masking strategy during training, obscuring 70--100\% of the audio latent. This design enforces consistency between the training objective (``context + mask'') and inference-time workflows (``reference audio + generated audio''). Both during training and inference, an \textit{InputEmbedding} layer processes the concatenated masked latent and intermediate state $X_t$ to inject conditional cues. To ensure compatibility with tasks where no reference audio is provided, the masked latent is zeroed out with a 30\% probability prior to concatenation. The resulting features are enriched via a \textit{ConvPositionEmbedding} module \cite{voicebox} for sequence positional encoding before being projected into the MMDiT input space. At inference, the conditional latent is injected through the \textit{InputEmbedding} layer and summed with initial noise, rendering the segment equivalent to the training-time masked region. To align audio and video durations, a surrogate video is prepended to the target video; potential misalignments are mitigated via synchronized masking on the video branch.

\noindent\textbf{Text modality} Semantic features are extracted via a CLIP \cite{clip} text encoder. Given that textual descriptors do not maintain a strictly frame-to-frame temporal alignment with audio, masking is not applied to the text modality.

\noindent\textbf{Video modality} Visual semantic features are derived from a CLIP visual encoder, while temporal synchronization features are extracted using Synchformer. To neutralize the influence of potentially misaligned features from the surrogate reference video introduced during inference, visual semantics undergo bilinear mapping-based masking, precisely mirroring the mask applied to the audio latents. Since synchronization features are projected and upsampled for temporal alignment, we investigated an alternative approach of similarly upsampling visual semantic features prior to masking. However, this resulted in suboptimal performance as upsampling-induced padding artifacts diluted the original semantic density. After undergoing an identical masking process, synchronization features are combined with the semantic features, which have been average-pooled and projected via a Multi-Layer Perceptron (MLP), to ensure that unsynchronized reference segments do not disrupt the alignment between the generated audio and the target video.

\noindent\textbf{Conditioning}
Multi-modal fusion is achieved by concatenating projected video and text semantic features with audio latents for MMDiT processing. Simultaneously, these semantic features are average-pooled and projected via an MLP to generate a global conditioning vector. This vector is combined with visual synchronization features to produce a frame-aligned condition. Both conditions are injected into the MMDiT blocks through the adaLN mechanism, with the frame-aligned condition further guiding the single-modal DiT.

\noindent\textbf{Loss function design}
We implement a Masked Mean Squared Error (MSE) loss, $ L_{\text{MSE}_\text{Masked}} $, which restricts gradient descent to the masked regions of each sample. While the baseline MMAudio loss, $L_{\mathrm{MSE}_{\mathrm{MMAudio}}}$, provides a global average over batch (B), time (T), and features (F), our adaptation first computes a per-frame loss:
\begin{align}
L_{\mathrm{MSE}_{\mathrm{Frame}}(b,t)} = \frac{1}{\mathrm{F}} \sum_{\mathrm{f}=1}^{\mathrm{F}} \left( \hat{\mathrm{y}}_{\mathrm{b},\mathrm{t},\mathrm{f}} - \mathrm{y}_{\mathrm{b},\mathrm{t},\mathrm{f}} \right)^2.
\end{align}
Using a boolean indicator $ \mathrm{rand\_span\_mask}(b,t) $, we isolate the masked frames. The final objective is normalized by the total number of masked frames $N$:
\begin{equation}
\begin{split}
L_{\mathrm{MSE}_{\mathrm{Masked}}} &= \frac{1}{N} \sum_{\mathrm{b}=1}^{\mathrm{B}} \sum_{\mathrm{t}=1}^{\mathrm{T}} \bigl( L_{\mathrm{MSE}_{\mathrm{Frame}}}(\mathrm{b,t}) \\
&\quad \times \mathrm{rand\_span\_mask}(\mathrm{b,t}) \bigr),
\end{split}
\label{eq:mse_masked}
\end{equation}
where $ N = \sum_{\mathrm{b}=1}^{\mathrm{B}} \sum_{\mathrm{t}=1}^{\mathrm{T}} \mathrm{rand\_span\_mask}(b,t) $.

\subsubsection{Distinction from MMAudio architecture}
FoleyGenEx represents a structural evolution over the MMAudio framework through several critical innovations:
\begin{itemize}
\item \textbf{Conditional injection pathway:} Unlike MMAudio, which lacks reference audio conditioning, FoleyGenEx introduces a dedicated injection mechanism. This pathway facilitates the integration of conditional latents with flow states via channel-wise concatenation and residual summation during inference, enabling sophisticated tasks like style transfer.
\item \textbf{Multimodal disentanglement via masking:} We implement a comprehensive multimodal masking strategy across audio, visual, and synchronization streams. This ensures train-inference consistency and prevents the model from developing shortcut biases between unsynchronized modalities.
\item \textbf{Segment-focused optimization:} By replacing global MSE with a masked MSE loss, FoleyGenEx prioritizes the reconstruction of precisely aligned segments, leading to superior temporal synchronization.
\end{itemize}

\begin{figure*}[t]
\setlength{\abovecaptionskip}{0.1cm} 
\centerline{\includegraphics[width=15cm]{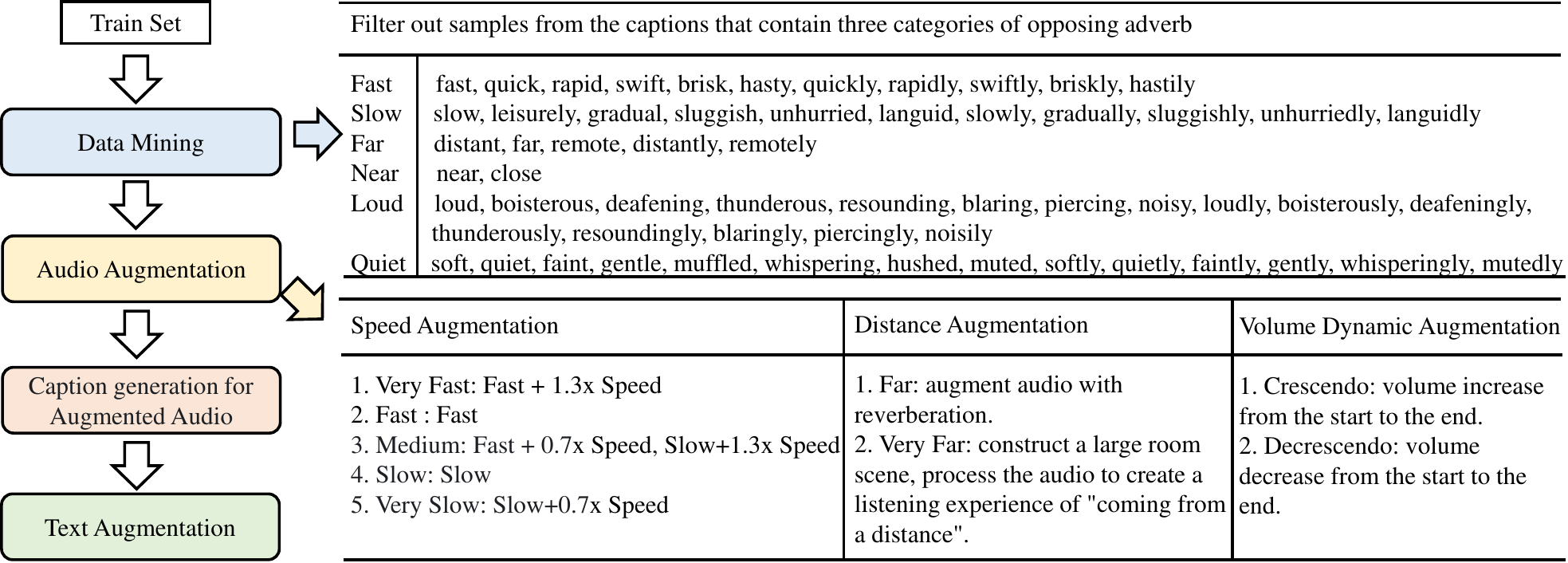}}
\caption{Adverb-based data augmentation algorithm.}
\label{f1:advaug}
\vspace{-20pt}
\end{figure*}

\subsection{Multi-modal-controlled audio generation}
\label{sec:multimodal-c}
The flexible adaptation of the FoleyGenEx architecture to diverse input modalities enables five distinct paradigms of multi-modal controlled audio generation, as illustrated in Fig. \ref{f3:inference-setup}. These tasks include Text-to-Audio (TTA), Video-to-Audio (VTA), Text-Controlled VTA (TC-VTA), Audio-Controlled VTA (AC-VTA), and Foley Extension (FE). Furthermore, by leveraging the latent inversion property \cite{latent-inversion} of the DiT backbone, we introduce a sixth capability: fine-grained audio editing through precise temporal segment control.

\noindent\textbf{TTA}
In the TTA setting, only text semantic features are provided, while all other modality streams are initialized as all-zero vectors. This configuration guides the model to synthesize audio that is strictly aligned with the linguistic semantics of the input prompt.

\noindent\textbf{VTA}
Both semantic and synchronization features of the target video are supplied. Textual input remains optional; however, if provided, it must be semantically consistent with the video content to avoid modal conflicts. This setup ensures that the generated audio is both temporally synchronized with visual events and aligned with the combined semantics derived from the video content and the complementary information provided by the text.

\noindent\textbf{TC-VTA}
By nullifying video semantic features (setting them to zero), the generation is guided by a combination of text semantics and video synchronization cues. Crucially, this allows for semantic-visual decoupling, where the text content need not match the video. For instance, a video of a cat meowing paired with the prompt ``Lion roaring" can synthesize a roar that is perfectly synchronized with the cat's mouth movements.

\noindent\textbf{AC-VTA}
In this task, both video and text semantic features are set to zero (F\textsubscript{V0}, F\textsubscript{T0}). As detailed in the inference setup illustrated in Figure \ref{f2:ac-vta-infer-set}, the reference audio latent is integrated through both channel-wise concatenation and residual summation with the initial noise. To maintain temporal structure, a surrogate reference video is constructed by cropping or duplicating segments from the target video to match the reference audio's duration. Its synchronization features (F\textsubscript{RS}) are concatenated prior to the target video's features (F\textsubscript{S}). This paradigm utilizes the timbre, prosody, and acoustic events of the reference audio as semantic anchors, enabling cross-modal style transfer. A practical application includes generating metal-knocking sounds synchronized with vegetable-chopping actions using a metal-knocking clip as a reference.

\noindent\textbf{FE}
Target video semantic and synchronization features are provided alongside a reference latent extracted from an existing audio segment. This latent is summed with the initial noise to provide conditional guidance, ensuring that the generated audio maintains stylistic and temporal continuity with the original segment.

\noindent\textbf{Editing}
Building upon the TC-VTA, AC-VTA, and FE frameworks, we utilize latent inversion to facilitate fine-grained audio manipulation. This enables the regeneration of audio within specific temporal windows, which are then seamlessly integrated with unedited regions in the latent space. By operating within the latent domain rather than the waveform level, our method avoids the audible clipping artifacts typically associated with standard concatenation. Qualitative results of this local editing feature are available on our project page\footnote{\url{https://foleygenex.github.io/FoleyGenEx}}.

\begin{figure*}[t]
\setlength{\abovecaptionskip}{0.1cm} 
\centerline{\includegraphics[width=14cm]{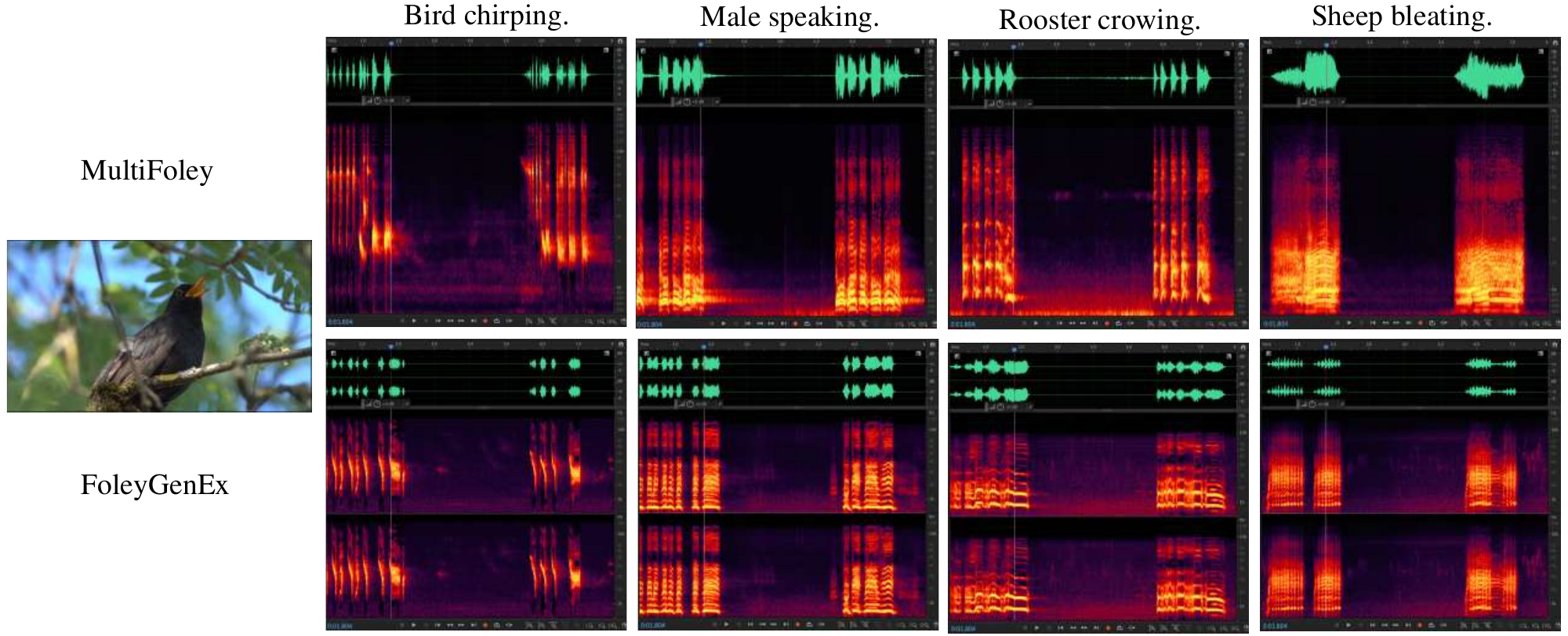}}
\caption{Spectrograms of audio generated by MultiFoley and our FoleyGenEx.}
\label{f3:tc-v2a-compa-multifoley}
\vspace{-10pt}
\end{figure*}

\subsection{Adverb-based data augmentation}
\label{sec:adverb-aug}
While architectural advancements expand the functional scope of multi-modal audio generation, the role of high-quality, fine-grained data remains paramount for achieving precise control. As established in Section \ref{sec:multimodal-c}, text serves as the primary semantic controller ; however, existing models predominantly prioritize nouns and actions, often overlooking the nuanced physical attributes conveyed by adverbs (e.g., speed, distance, and volume). We attribute this deficiency to the scarcity of adverb-rich training samples in standard benchmarks and the prohibitive costs associated with manual annotation. To bridge this gap and enable refined adverbial control, we propose an automated, four-stage data augmentation algorithm (Fig. \ref{f1:advaug}):

\noindent\textbf{Data mining}
We focus on three orthogonal dimensions of opposing adverbs: Speed (\textit{Fast/Slow}), Distance (\textit{Far/Near}), and Volume (\textit{Loud/Quiet}). A base dataset is curated by filtering training samples that explicitly contain these adverbial categories or their corresponding adjective forms.

\noindent\textbf{Audio augmentation}
Targeted acoustic modifications are applied to the base dataset to synthesize diverse physical variations:
\begin{itemize}
\item \textbf{Speed augmentation}: For speed-related samples, we construct a five-level scale by modulating audio playback rates: (1) \textit{Very Fast}: Original `Fast' samples increased by 1.3$\times$ speed. (2) \textit{Fast}: Original `Fast' samples. (3) \textit{Medium}: Original `Fast' samples slowed by 0.7$\times$ or `Slow' samples increased by 1.3$\times$. (4) \textit{Slow}: Original `Slow' samples. (5) \textit{Very Slow}: Original `Slow' samples reduced by 0.7$\times$ speed.
\item \textbf{Distance augmentation}: For `Near' or `Loud' samples, which possess high initial signal energy, we apply distance simulation by introducing room reverberation and spectral attenuation.
\item \textbf{Volume dynamic augmentation}: For `Loud' samples, we modulate the amplitude envelope to create dynamic trends—such as \textit{Crescendo} (increasing volume) and \textit{Decrescendo} (decreasing volume)—while preserving the original peak amplitude.
\end{itemize}

\noindent\textbf{Caption generation for augmented audio}
To ensure semantic alignment, we employ LLMs with specialized prompts—the details of which are provided on our project page—to generate new captions for the augmented audio. These prompts synthesize the original context with the specific physical modification applied. 

\noindent\textbf{Text augmentation}
To bolster linguistic generalization, LLMs are used to extract key semantic tokens and produce paraphrased versions of the augmented captions. This strategy accounts for the inherent variability in natural language, where a single acoustic event can be described through diverse syntactic structures.

We conducted a stratified manual verification on 300 randomly sampled entries to ensure audio-caption correspondence. The set included 20 samples per speed level and 50 samples for each distance and volume operation type. This verification process yielded an accuracy rate exceeding 97\%, confirming the reliability of our augmented dataset.

\begin{figure*}[t]
\setlength{\abovecaptionskip}{0.1cm} 
\centerline{\includegraphics[width=14cm]{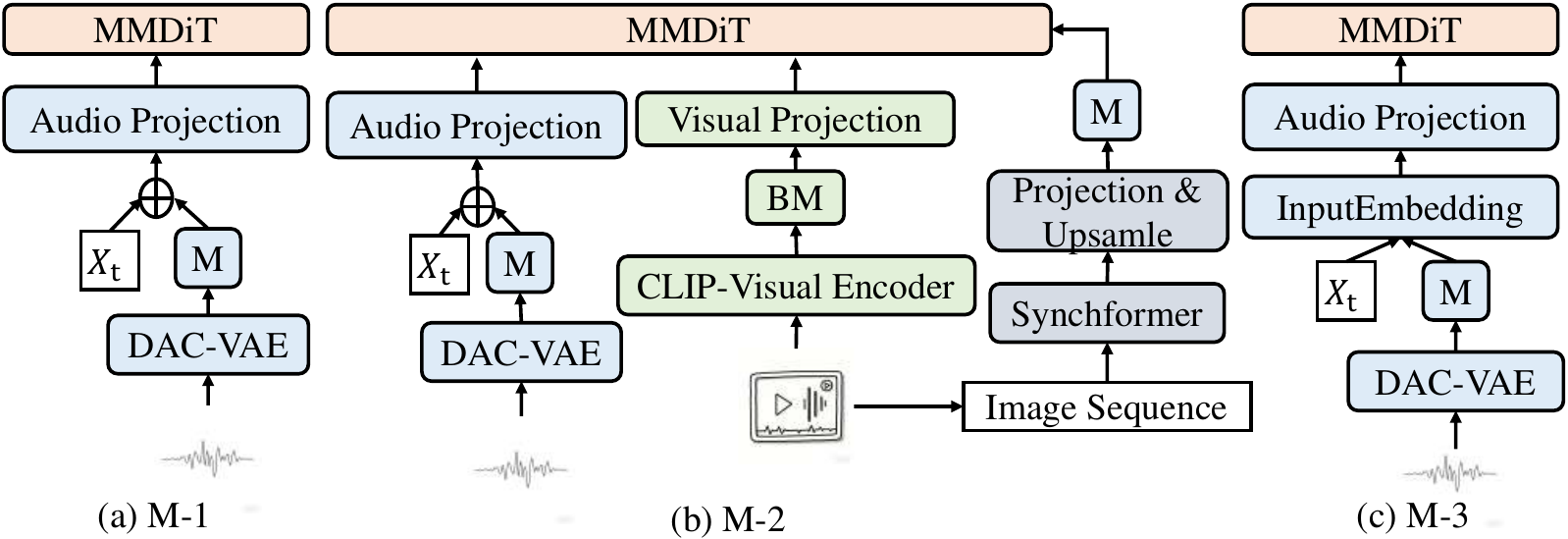}}
\caption{Different mask training strategies: (a) Masking on the audio branch (b) Multimodal masking (c) Masking on the audio branch with InputEmbedding module. }
\label{f1:diff-train-mask}
\vspace{-15pt}
\end{figure*}

\setlength{\tabcolsep}{0.8mm}{
\begin{table}[t]
\setlength{\abovecaptionskip}{0.0cm} 
\caption{Performance on the AudioCaps test set. The results of all baseline are from \cite{MMAudio}. Bold font indicates the optimal performance under each metric, while underlined text marks the second-best performance.}
\vspace{-10pt}
\begin{center}
\begin{tabular}{l c c c }
\hline
\textbf{Method} & \textbf{FD\textsubscript{VGG}} $\downarrow$ & \textbf{IS} $\uparrow$ & \textbf{CLAP\textsubscript{T}} $\uparrow$ \\
\hline
GenAU-Large \cite{GenAU-Large} & \textbf{1.21} & 11.75 & 0.285 \\
MMAudio \cite{MMAudio}& 4.03 & \textbf{12.08} & 0.348\\
FoleyGenEx (Ours) & 2.61 & 11.80 & \underline{0.364}\\
FoleyGenEx + AA (Ours) & 2.60 & \underline{11.89} & \textbf{0.366} \\
\hline
\end{tabular}
\label{t3:audiocaps}
\end{center}
\vspace{-18pt}
\end{table}
}

\setlength{\tabcolsep}{0mm}{
\begin{table}[t]
\setlength{\abovecaptionskip}{0.0cm} 
\caption{Performance on the VGGSound test set.}
\vspace{-10pt}
\begin{center}
\begin{tabular}{l c c c c} 
\hline
\textbf{Model} & \textbf{FD\textsubscript{VGG}} $\downarrow$ & \textbf{IS} $\uparrow$ & \textbf{IB-score} $\uparrow$ & \textbf{DeSync} $\downarrow$ \\
\hline
VTA-LDM \cite{vta-ldm} & 2.04 & 10.14 & 24.72 & 1.263\\
FoleyCrafter \cite{foleycrafter} & 2.51 & 15.68 & 25.68 & 1.225 \\
MMAudio \cite{MMAudio} & 0.97 & 17.40 & \textbf{33.22} & \textbf{0.442} \\ 
FoleyGenEx (Ours) & \underline{0.74} & \underline{18.37} & 32.52 & 0.467\\
FoleyGenEx + AA (Ours) & \textbf{0.73} & \textbf{18.49} & \underline{33.17} & \underline{0.453} \\
\hline
MultiFoley * \cite{multifoley} & 2.92 & --- & 28.00 & 0.800 \\ 
VTA-LDM * & 2.02 & 11.32 & 28.34 & 1.275\\
FoleyCrafter * & 2.74 & 16.15 & 30.20 & 1.240 \\ 
MMAudio * & 1.13 & 17.59 & \underline{37.85} & \textbf{0.393} \\
FoleyGenEx (Ours) * & \underline{0.87} & \underline{18.64} & 37.23 & 0.409 \\
FoleyGenEx + AA (Ours) * & \textbf{0.86} & \textbf{19.56} & \textbf{38.06} & \underline{0.403}\\
\hline
\end{tabular}
\label{t2:vggsound}
\end{center}
\vspace{-30pt}
\end{table}
}

\begin{table}[t]
\setlength{\abovecaptionskip}{0cm} 
\caption{Comparison of text-controlled video-to audio performance.}
\vspace{-10pt}
\begin{center}
\begin{tabular}{l c c}
\hline
\textbf{Method} & \textbf{DeSync} $\downarrow$ & \textbf{CLAP\textsubscript{T}} $\uparrow$ \\
\hline
MultiFoley \cite{multifoley} & 0.169 & 31.39\\
MMAudio \cite{MMAudio} & \textbf{0.053} & 32.53\\
FoleyGenEx (Ours) & 0.061 & \underline{33.53} \\
FoleyGenEx + AA (Ours) & \underline{0.059} & \textbf{34.20} \\
\hline
\end{tabular}
\label{t4:tc-v2a}
\end{center}
\vspace{-28pt} 
\end{table}

\section{Experimental settings}
\label{sec:experimental-settings}
\subsection{Datasets}
\label{sec:train-dataset}
FoleyGenEx is trained on a composite dataset encompassing both VTA and TTA tasks, maintaining a base configuration consistent with MMAudio. To evaluate the impact of fine-grained semantic control, specific training runs incorporate our custom adverb-augmented data, denoted by the `AA' identifier when included in the TTA subset. The dataset composition is detailed as follows:
\begin{itemize}
\item \textbf{VTA data}: Approximately 500 hours of audio-visual content sourced from VGGSound. All videos are truncated to 8-second segments, specifically retaining the final 8 seconds to exclude irrelevant introductory content. The dataset's native category labels serve directly as the textual supervision.
\item \textbf{TTA base data}: Comprises approximately 128 hours from AudioCaps and 7,600 hours from WavCaps\cite{wavcaps}.
\begin{itemize}
\item \textbf{AudioCaps}: Audio clips are uniformly truncated to 8-second segments.
\item \textbf{WavCaps}: After filtering invalid text samples, clips shorter than 16 seconds are truncated to 8 seconds. Clips exceeding 16 seconds are segmented into up to five non-overlapping 8-second portions to maximize data utilization while minimizing redundancy.
\end{itemize}
\item \textbf{Adverb-augmented (AA) data}: Consists of 88,370 in-house samples systematically constructed through a multi-stage pipeline. Audio augmentations are precisely guided by adverbial cues extracted from original captions:
\begin{enumerate}
\item \textbf{Silence removal}: FFmpeg is employed to eliminate leading and trailing silence from all raw audio samples.
\item \textbf{Targeted augmentation}: Specialized tools facilitate distinct physical modifications: SoX for speed adjustment, Pyroomacoustics for distance simulation (via room reverberation), and FFmpeg for gradual dynamic volume scaling.
\item \textbf{Caption refinement}: Corresponding captions for the augmented audio are generated and optimized using LLMs, with prompts available on our project page.
\end{enumerate}
\end{itemize}

\subsection{Implementation details}
\label{sec:training-and-evaluation-setup}
Our architecture and training procedure are implemented based on the MMAudio framework\footnote{\url{https://github.com/hkchengrex/MMAudio}}. We adopt the learning rate, scheduler, and MMAudio-L-44.1kHz hyperparameter configuration from the original settings, utilizing a batch size of 256. Base training was conducted for 300,000 steps, requiring approximately three days on a machine equipped with 8 A100 GPUs. When incorporating the AA data, training was extended to 330,000 steps to maintain a comparable data traversal. 
During inference, FoleyGenEx utilizes a classifier-free guidance scale of 4.5 and 25 flow-matching steps to generate the audio samples.
To facilitate Classifier-Free Guidance (CFG) \cite{cfg} during inference, we randomly drop either video or text features with a 10\% probability. Efficiency and balanced learning are ensured through two key strategies:
\begin{itemize}
\item \textbf{Feature precomputation}: Audio latents, visual features, and text embeddings are precomputed and stored as binary files to eliminate on-the-fly computation overhead.
\item \textbf{Balanced sampling}: VTA and TTA samples are processed at a strict 1:1 ratio within each training batch to prevent biased learning trajectories.
\end{itemize}

Performance is evaluated across four critical dimensions using the \textit{av-benchmark} toolkit\footnote{\url{https://github.com/hkchengrex/av-benchmark}}:
\begin{itemize}
\item \textbf{Distribution matching}: Assessed via Fréchet Distance (FD) between generated and real features using VGGish \cite{vggish} (FD\textsubscript{VGG}).
\item \textbf{Audio quality}: Evaluated using the Inception Score (IS) \cite{inceptionscore} with PANNs \cite{panns}.
\item \textbf{Semantic alignment}: Quantified by the cosine similarity between embeddings extracted from different modalities:
\begin{itemize}
\item For TTA: CLAP\textsubscript{T} measures the similarity between text features and audio features, both extracted using the respective encoders of the CLAP model \cite{clap}.
\item For VTA: IB-score measures the similarity between visual features and audio features, extracted via the ImageBind \cite{imagebind} multi-modal encoders.
\end{itemize}
\item \textbf{Temporal alignment}: Quantified by the audio-video misalignment in seconds as predicted by Synchformer (DeSync).
\end{itemize}

Furthermore, to assess semantic consistency regarding adverbs, we conducted a Good, Same, Bad (GSB) subjective study. Ten evaluators compared model outputs in a blinded A/B preference test. 
Evaluators assessed the MMAudio model retrained with our AA data against the original pre-trained MMAudio baseline based on semantic consistency with adverbial cues.
The presentation order was randomized to mitigate bias across 500 total assessments (10 evaluators $\times$ 50 samples). The final score is calculated as: $Score = (G + S) / (S + B)$.

\setlength{\tabcolsep}{0.8mm}{
\begin{table*}[t]
\setlength{\abovecaptionskip}{0.0cm} 
\caption{Comparison of audio-controlled video-to audio performance.}
\vspace{-10pt}
\begin{center}
\begin{tabular}{l c c c c}
\hline
\textbf{Method} & \textbf{OnsetSyncAP(\%)} $\uparrow$ & \textbf{FD\textsubscript{VGG}} $\downarrow$ & \textbf{Resemblyzer} $\uparrow$ & \textbf{CLAP\textsubscript{A}} $\uparrow$  \\
\hline
CondFoleyGen \cite{CondFoleyGen} & 60.00 & \underline{0.65} & 0.8999 & \textbf{0.7385} \\
MMAudio-S1 & 65.53 & 4.77 & 0.8185 & 0.5063 \\
MMAudio-S2 & 68.72 & 3.05 & 0.8360 & 0.5132\\
MMAudio-S3 & 68.92 & 2.16 & 0.8568 & 0.5195 \\
FoleyGenEx (Ours) & \underline{69.38} & \textbf{0.54} & \underline{0.9085} & 0.7216\\
FoleyGenEx + AA (Ours) & \textbf{69.71} & \textbf{0.54} & \textbf{0.9128} & \underline{0.7250} \\
\hline
\end{tabular}
\label{t5:ac-v2a}
\end{center}
\vspace{-18pt}
\end{table*}
}

\setlength{\tabcolsep}{0.5mm}{
\begin{table*}[t]
\setlength{\abovecaptionskip}{0.0cm} 
\caption{Comparison of Foley extension performance. C\textsubscript{A}:CLAP\textsubscript{A}, DS:DeSync. }
\vspace{-10pt}
\begin{center}
\begin{tabular}{c | c c c c | c c | c c | c c | c c | c c | c c | c c}
\hline
\textbf{Sample} & \multicolumn{4}{c|}{\textbf{Conditions}} &  \multicolumn{2}{c|}{\textbf{MultiFoley}} & \multicolumn{2}{c|}{\textbf{MMAudio}} & \multicolumn{2}{c|}{\textbf{FoleyGenEx}} & \multicolumn{2}{c|}{\textbf{FoleyGenEx + AA}} & \multicolumn{2}{c|}{\textbf{M-1}} & \multicolumn{2}{c|}{\textbf{M-2}} & \multicolumn{2}{c}{\textbf{M-3}} \\  
\textbf{Size} & \textbf{V\textsubscript{t}} & \textbf{Text} & \textbf{A\textsubscript{r}} & \textbf{V\textsubscript{r}} & \textbf{C\textsubscript{A}} $\uparrow$ & \textbf{DS} $\downarrow$ & \textbf{C\textsubscript{A}} $\uparrow$ & \textbf{DS} $\downarrow$ & \textbf{C\textsubscript{A}} $\uparrow$ & \textbf{DS} $\downarrow$ & \textbf{C\textsubscript{A}} $\uparrow$ & \textbf{DS} $\downarrow$ & \textbf{C\textsubscript{A}} $\uparrow$ & \textbf{DS} $\downarrow$ & \textbf{C\textsubscript{A}} $\uparrow$ & \textbf{DS} $\downarrow$ & \textbf{C\textsubscript{A}} $\uparrow$ & \textbf{DS} $\downarrow$ \\
\hline
& \checkmark & \checkmark & & & 55.4 & 0.79 & 56.0 & 0.40 & \underline{59.7} & \underline{0.39} & \textbf{60.5} & \textbf{0.38} & 57.5 & 0.40 & 57.5 & 0.40 & 58.9 & 0.39 \\
\textbf{1000} & \checkmark & & \checkmark & & 59.6 & 0.78 & 57.0 & \underline{0.46} & \underline{61.8} & \textbf{0.40} & \textbf{62.3} & \textbf{0.40} & 58.1 & 0.46 & 58.2 & 0.44 & 60.9 & 0.42 \\
& \checkmark & & \checkmark & \checkmark & 59.8 & 0.77 & 59.0 & 0.39 & \underline{69.3} & \textbf{0.37} & \textbf{69.7} & \textbf{0.36} & 65.9 & 0.39 & 66.2 & 0.38 & 68.7 & 0.38 \\
& \checkmark & \checkmark & \checkmark & \checkmark & 64.3 & 0.77 & 60.7 & \underline{0.38} & \underline{71.2} & \textbf{0.36} & \textbf{71.5} & \textbf{0.36} & 70.0 & 0.37 & 70.2 & 0.37 & 70.9 & 0.36 \\
\hline
& \checkmark & \checkmark & & & --- & --- & 50.7 & \underline{0.41} & \underline{51.6} & \textbf{0.39} & \textbf{51.8} & \textbf{0.39} & --- & --- & --- & --- & --- & --- \\
\textbf{ALL} & \checkmark & & \checkmark & & --- & --- & 52.8 & 0.47 & \underline{62.7} & \underline{0.40} & \textbf{62.9} & \textbf{0.39} & --- & --- & --- & --- & --- & --- \\
& \checkmark & & \checkmark & \checkmark & --- & ---& 53.6 & 0.39 & \underline{63.5} & \underline{0.38} & \textbf{63.7} & \textbf{0.37} & --- & --- & --- & --- & --- & --- \\
& \checkmark & \checkmark & \checkmark & \checkmark & --- & --- & 57.4 & \underline{0.38} & \underline{65.8} & \textbf{0.36} & \textbf{65.9} & \textbf{0.36} & --- & --- & --- & --- & --- & --- \\
\hline
\end{tabular}
\label{t6:fe}
\end{center}
\vspace{-24pt}
\end{table*}
}

\begin{figure*}[t]
\setlength{\abovecaptionskip}{0.1cm} 
\centerline{\includegraphics[width=13.5cm]{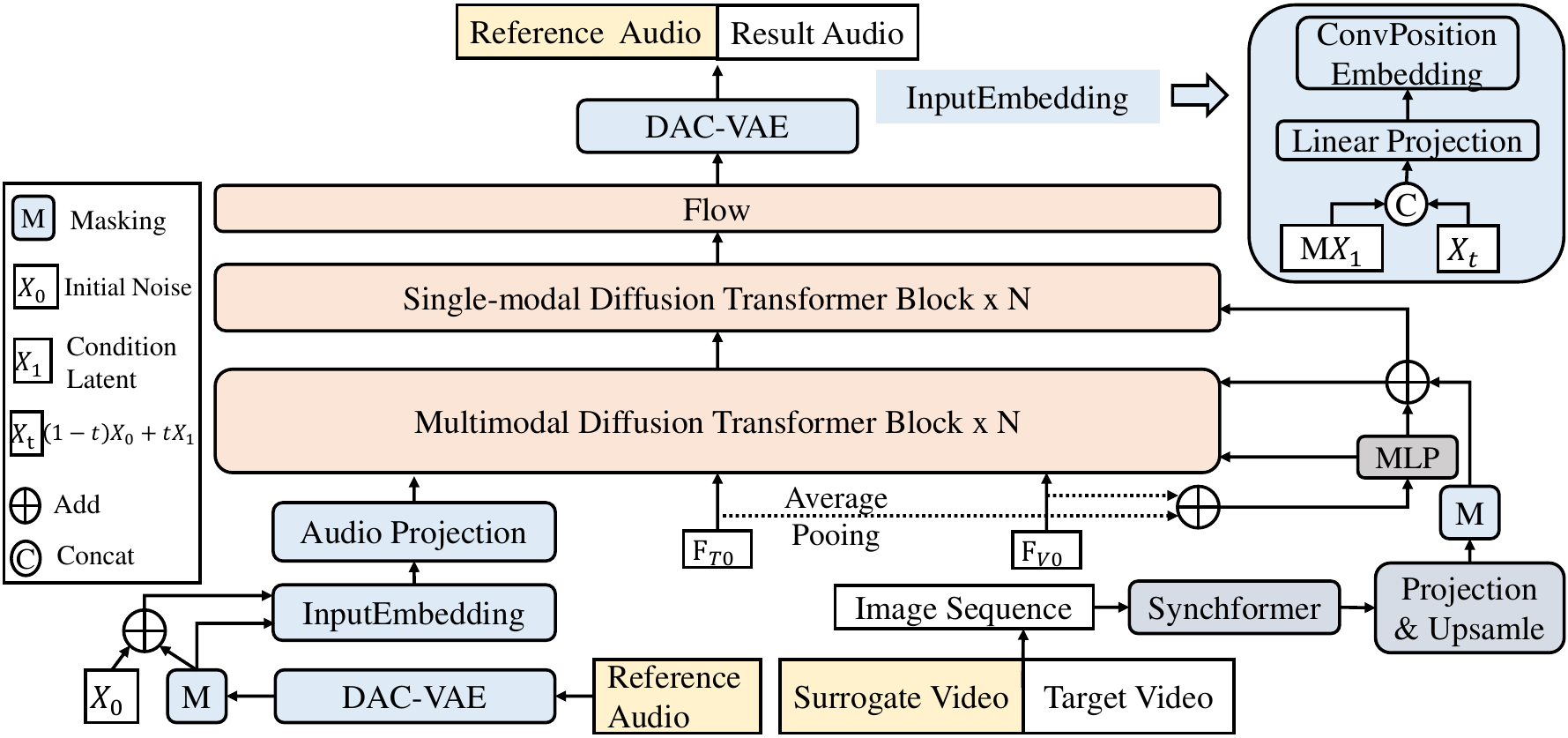}}
\caption{The inference setup for the audio-controlled video-to-audio task.}
\label{f2:ac-vta-infer-set}
\vspace{-10pt}
\end{figure*}

\setlength{\tabcolsep}{0mm}{
\begin{table}[t]
\setlength{\abovecaptionskip}{0cm} 
\caption{Performance of adverb-augmented data added to the training set.}
\vspace{-10pt}
\begin{center}
\begin{tabular}{l c c c c c c}
\hline
\textbf{Setting} & \textbf{FD\textsubscript{VGG}} $\downarrow$ & \textbf{KL\textsubscript{PANNS}} $\downarrow$ & \textbf{KL\textsubscript{PASST}} $\downarrow$ & \textbf{CLAP\textsubscript{T}} $\uparrow$ \\
\hline
MMAudio & 7.67 & 1.59 & 1.67 & 0.323\\
FoleyGenEx (Ours) & 3.07 & 1.45 & 1.49 & 0.354 \\
\hline
MMAudio+AA & 4.03 & 1.47 & 1.45 & 0.361\\
FoleyGenEx (Ours)+AA & \textbf{2.96} & \textbf{1.41} & \textbf{1.43} & \textbf{0.365} \\
\hline
\end{tabular}
\label{t7:adv-aug}
\end{center}
\vspace{-28pt}
\end{table}
}

\setlength{\tabcolsep}{0.8mm}{
\begin{table*}[t]
\setlength{\abovecaptionskip}{0cm} 
\caption{Comparison of audio-controlled video-to-audio performance across different mask training strategies.}
\vspace{-10pt}
\begin{center}
\begin{tabular}{l c c c c}
\hline
\textbf{Method} & \textbf{OnsetSyncAP(\%)} $\uparrow$ & \textbf{FD\textsubscript{VGG}} $\downarrow$ & \textbf{Resemblyzer} $\uparrow$ & \textbf{CLAP\textsubscript{A}} $\uparrow$  \\
\hline
M-1 & 66.97 & 3.45 & 0.8430 & 0.5141 \\ %
M-2 & 68.90 & 2.91 & 0.8470 & 0.5224 \\ %
M-3 & 67.95 & 3.13 & \underline{0.8677} & \underline{0.7202} \\ %
FoleyGenEx (w/o summation) & \underline{69.26} & \underline{0.65} & 0.8562 & 0.7106 \\
FoleyGenEx & \textbf{69.38} & \textbf{0.54} & \textbf{0.9085} & \textbf{0.7216} \\
\hline
\end{tabular}
\label{t1:ac-v2a}
\end{center}
\vspace{-24pt}
\end{table*}
}

\section{Results and discussions}
\label{sec:results}
\subsection{TTA \& VTA performance}
\label{sec:tta-vta}
We evaluated the TTA task on the AudioCaps test set (964 samples) and the VTA task on the VGGSound test set (15,220 samples). To ensure a rigorous comparison with MultiFoley, which excludes VGGSound samples with an IB-score below 0.3, we applied the same filter to our test set, resulting in 8,702 samples. In Table \ref{t2:vggsound}, results marked with `*' correspond to this filtered subset, while others refer to the complete set. We reproduced inference results for MMAudio, FoleyCrafter\footnote{\url{https://github.com/open-mmlab/FoleyCrafter}}, and VTA-LDM\footnote{\url{https://github.com/ariesssxu/vta-ldm}} using their open-source implementations for a balanced comparison. All models generated 8-second clips.

As detailed in Tables \ref{t3:audiocaps} and \ref{t2:vggsound}, FoleyGenEx outperforms MMAudio in distribution matching (FD\textsubscript{VGG}) and semantic relevance (CLAP\textsubscript{T}) for TTA while maintaining competitive audio quality. For VTA, our model shows consistent improvements over MMAudio in distribution matching and audio quality, alongside competitive temporal alignment. These metrics demonstrate that our conditional injection and multimodal dynamic masking strategies effectively extend the framework’s versatility without compromising its fundamental generative performance. Furthermore, our method exhibits significant advantages across all dimensions compared to MultiFoley.

\subsection{TC-VTA: semantic-synchronization decoupling}
\label{sec:tc-vta}
In the TC-VTA task, where input text is semantically independent of the video, we nullified video semantic features to test the model's ability to handle modal conflicts. Using 13 examples from the MultiFoley benchmark\footnote{\url{https://ificl.github.io/MultiFoley/}}, we compared semantic relevance (CLAP\textsubscript{T}) and temporal alignment (DeSync).

Table \ref{t4:tc-v2a} shows that FoleyGenEx significantly outperforms MultiFoley in both metrics. This is attributed to the MMDiT architecture’s refined processing of semantic and synchronization features separately, whereas MultiFoley’s simplistic upsampling often leads to degraded alignment. Spectrogram analysis (Fig. \ref{f3:tc-v2a-compa-multifoley}) confirms this: at 1.804s (a bird opening its beak), MultiFoley only generates vocalization spectral features if the text matches the action. In contrast, FoleyGenEx accurately captures the visual action, generating precisely timed audio regardless of the textual input.

\subsection{AC-VTA: style transfer and zero-shot generalization}
\label{sec:ac-vta}
For the AC-VTA task, we adopted the evaluation protocol from CondFoleyGen\footnote{\url{https://github.com/XYPB/CondFoleyGen/tree/main}}, utilizing a test set derived from the Greatest Hits dataset. This setup involves generating 582 audio samples by performing style transfer on 194 two-second target video clips, each paired with three distinct reference audio samples. Since the original MMAudio framework lacks a native reference audio conditioning branch, we adapted the model by concatenating the reference audio latents with the initial noise and intermediate states $X_t$ to provide stylistic guidance. Due to the inherent nature of the MMDiT architecture—which generates audio with a duration strictly matching the input video sequence—a surrogate reference video is prepended to the target video to extend the total input length.
We explored three surrogate video configurations, where a two-second segment is prepended to the target video to match the reference audio's duration:
\begin{itemize}
    \item \textbf{MMAudio-S1}: A two-second blank (zero-filled) video is prepended.
    \item \textbf{MMAudio-S2}: The first two seconds of the target video are duplicated and prepended as a visual placeholder.
    \item \textbf{MMAudio-S3}: The actual reference video corresponding to the reference audio is prepended.
\end{itemize}

Performance was assessed across multiple dimensions: temporal alignment via OnsetSyncAP (following the evaluation protocol in CondFoleyGen), distribution matching via FD\textsubscript{VGG}, style similarity (timbre and prosody) via a Resemblyzer-based\footnote{\url{https://github.com/resemble-ai/Resemblyzer}} metric, and semantic consistency via CLAP\textsubscript{A} (cosine similarity between reference and generated audio embeddings). 

As demonstrated in Table \ref{t5:ac-v2a}, FoleyGenEx achieves robust performance across all metrics. While our CLAP\textsubscript{A} score is slightly lower than CondFoleyGen's, it is critical to note that the CondFoleyGen model evaluated here was explicitly trained on in-domain Greatest Hits data. In contrast, FoleyGenEx achieves highly competitive results through zero-shot generalization. Furthermore, although the MMAudio-S3 configuration showed commendable temporal alignment by leveraging the actual reference video, it failed to capture the stylistic nuances of the reference audio, as evidenced by its significantly lower CLAP\textsubscript{A} score. 
This discrepancy validates that our specialized conditional injection and multimodal masking strategies are essential for effective reference-based conditioning.

\subsection{FE: style and temporal continuity}
\label{sec:fe}
For the FE task, we followed the protocol established by MultiFoley, randomly selecting 1,000 eight-second clips from the VGGSound-Sync test set \cite{vggsound-sync} for primary evaluation. To ensure statistical robustness and eliminate sampling bias, we further report results on the complete test set, denoted as the `ALL' setting in Table \ref{t6:fe}. The objective was to generate synchronized audio for the final 5-second segment (3--8s), conditioned on the initial 3-second reference. We assessed performance across four distinct configurations derived from the same test clip, where each configuration is defined by the specific combination of input modalities provided. In Table 5, checkmarks indicate the specific conditional inputs supplied for each setting, including the target video (V\textsubscript{t}), semantic text (Text), reference audio (A\textsubscript{r}), and reference video (V\textsubscript{r}). 

As demonstrated in Table \ref{t6:fe}, FoleyGenEx significantly surpasses MultiFoley across all compared dimensions, particularly in temporal alignment (DeSync). A critical finding is that while MMAudio incorporates a robust synchronization module, it fails to maintain style continuity. This deficiency in MMAudio stems from the lack of a dedicated reference audio injection pathway and a multimodal masking strategy during training. Without these components, the model fails to learn how to leverage reference contexts effectively, leading to a severe discrepancy between training and inference.

\subsection{Adverb augmentation analysis}
\label{sec:adverb-exp}
To evaluate fine-grained textual control, we integrated the AA data into the training pipeline. We assess performance across two key dimensions: (1) Distribution matching, utilizing FD\textsubscript{VGG} along with KL\textsubscript{PANNS} and KL\textsubscript{PASST} (which compute the Kullback–Leibler distance between the distributions of generated and ground-truth audio using PANNs and PaSST \cite{passt} as embedding models); and (2) Semantic consistency, measured by CLAP\textsubscript{T} to ensure nuanced adverbial cues are accurately reflected. As shown in Table \ref{t7:adv-aug}, incorporating AA data leads to consistent enhancements across all metrics for both MMAudio and FoleyGenEx. 

To further validate the model's sensitivity to nuanced semantic cues, we manually curated a test set of 50 diverse adverb-containing captions (to be made publicly available). Following the setup in Section \ref{sec:experimental-settings}, ten evaluators compared audio generated by the original pre-trained MMAudio against the MMAudio model retrained with our AA data. The enhanced MMAudio achieved a decisive G:S:B ratio of 386:66:48, resulting in a superior subjective score of 3.965. These results indicate that in 77.2\% (386/500) of instances, evaluators found the AA-enhanced MMAudio outputs significantly more consistent with specific adverbial descriptors compared to the original baseline.

\subsection{Ablation experiment}
\label{sec:ablation}
We evaluated three distinct mask training strategies (Fig. \ref{f1:diff-train-mask}) to isolate the contributions of our proposed conditional injection and multimodal dynamic masking across the AC-VTA (Table \ref{t1:ac-v2a}) and FE tasks (Table \ref{t6:fe}). The configurations include: (a) M-1: audio-only branch masking; (b) M-2: multimodal masking across audio, visual, and synchronization streams; and (c) {M-3: audio-branch masking integrated with the \textit{InputEmbedding} module for reference conditioning.

\noindent\textbf{Multimodal masking and synchronization} 
For the AC-VTA task, multimodal masking (M-2) consistently outperformed audio-only strategies—even those equipped with the \textit{InputEmbedding} module—in temporal synchronization (OnsetSyncAP). This superiority is attributed to the multimodal strategy's ability to train the model to disentangle reference audio cues from surrogate visual features. Conversely, unimodal audio masking during training, where conditional audio is inherently aligned with its corresponding video segment, causes the model to develop a ``shortcut bias''. In such cases, the model incorrectly assumes this static alignment persists during inference when presented with randomly cropped target videos, leading to synchronization failure and performance inferior to the baseline MMAudio under identical test conditions (MMAudio-S2 in Table \ref{t5:ac-v2a}).

\noindent\textbf{Reference injection and style transfer} 
Regarding audio style transfer, all three strategies exhibited improvements in timbre and semantics over the original MMAudio-S2 benchmark. Crucially, the incorporation of the \textit{InputEmbedding} module (M-3 in Table \ref{t5:ac-v2a}) enabled FoleyGenEx to exceed the performance of MMAudio’s ideal configuration (MMAudio-S3, which utilizes true reference videos), significantly boosting semantic metrics such as CLAP\textsubscript{A}. 

\noindent\textbf{Impact on foley extension} 
In the FE task, temporal alignment remains inherently stable across all strategies due to the natural alignment of reference and target audio. However, the \textit{InputEmbedding} module still provided a noticeable enhancement in overall performance. Specifically, while strategies without this module already outperformed MMAudio, its integration—tailored for reference-based style transfer—further bolstered the model's ability to achieve seamless stylistic continuation between the reference segment and the generated content.

\section{Conclusions}
We introduces FoleyGenEx, an MMDiT-based VTA framework that addresses key limitations of existing methods by separating semantic and synchronization information processing. Equipped with a conditional injection mechanism for audio-guided tasks, a multi-modal dynamic masking strategy for stable temporal alignment, and an adverb-based data augmentation algorithm for fine-grained semantic control, FoleyGenEx unifies six core capabilities (TTA, VTA, TC-VTA, AC-VTA, FE, and audio editing). Experiments on multiple datasets confirm it retains MMAudio’s synchronization strength, extends MultiFoley’s control versatility, and adds unique adverb-level precision, outperforming baselines across critical metrics to advance controllable audio generation.


\section{Generative AI Use Disclosure}
During the preparation of this paper, LLMs were used as a tool for writing assistance and polishing. Specifically, LLMs helped refine language, improve clarity and fluency. All content generated or refined by LLMs was carefully reviewed and verified by the authors to ensure accuracy, originality, and compliance with academic integrity. The authors take full responsibility for the paper's final content, including any portions that utilized LLMs.

\bibliographystyle{IEEEtran}
\bibliography{mybib}

\begin{thebibliography}{10}
\providecommand{\url}[1]{#1}
\csname url@samestyle\endcsname
\providecommand{\newblock}{\relax}
\providecommand{\bibinfo}[2]{#2}
\providecommand{\BIBentrySTDinterwordspacing}{\spaceskip=0pt\relax}
\providecommand{\BIBentryALTinterwordstretchfactor}{4}
\providecommand{\BIBentryALTinterwordspacing}{\spaceskip=\fontdimen2\font plus
\BIBentryALTinterwordstretchfactor\fontdimen3\font minus \fontdimen4\font\relax}
\providecommand{\BIBforeignlanguage}[2]{{%
\expandafter\ifx\csname l@#1\endcsname\relax
\typeout{** WARNING: IEEEtran.bst: No hyphenation pattern has been}%
\typeout{** loaded for the language `#1'. Using the pattern for}%
\typeout{** the default language instead.}%
\else
\language=\csname l@#1\endcsname
\fi
#2}}
\providecommand{\BIBdecl}{\relax}
\BIBdecl

\bibitem{sora}
\BIBentryALTinterwordspacing
OpenAI, ``Video generation models as world simulators,'' 2024. [Online]. Available: \url{https://openai.com/index/video-generation-models-as-world-simulators/}
\BIBentrySTDinterwordspacing

\bibitem{cogvideox}
Z.~Yang, J.~Teng, W.~Zheng, M.~Ding, S.~Huang, J.~Xu, Y.~Yang, W.~Hong, X.~Zhang, G.~Feng \emph{et~al.}, ``Cogvideox: Text-to-video diffusion models with an expert transformer,'' \emph{arXiv preprint arXiv:2408.06072}, 2024.

\bibitem{hunyuanvideo}
W.~Kong, Q.~Tian, Z.~Zhang, R.~Min, Z.~Dai, J.~Zhou, J.~Xiong, X.~Li, B.~Wu, J.~Zhang \emph{et~al.}, ``Hunyuanvideo: A systematic framework for large video generative models, 2025,'' \emph{URL https://arxiv. org/abs/2412.03603}.

\bibitem{kling}
\BIBentryALTinterwordspacing
Kuaishou, ``Kling ai,'' 06 2024. [Online]. Available: \url{https://klingai.kuaishou.com/}
\BIBentrySTDinterwordspacing

\bibitem{wan}
T.~Wan, A.~Wang, B.~Ai, B.~Wen, C.~Mao, C.-W. Xie, D.~Chen, F.~Yu, H.~Zhao, J.~Yang \emph{et~al.}, ``Wan: Open and advanced large-scale video generative models,'' \emph{arXiv preprint arXiv:2503.20314}, 2025.

\bibitem{V-AURA}
I.~Viertola, V.~Iashin, and E.~Rahtu, ``Temporally aligned audio for video with autoregression,'' in \emph{ICASSP 2025-2025 IEEE International Conference on Acoustics, Speech and Signal Processing (ICASSP)}.\hskip 1em plus 0.5em minus 0.4em\relax IEEE, 2025, pp. 1--5.

\bibitem{VATT}
X.~Liu, K.~Su, and E.~Shlizerman, ``Tell what you hear from what you see-video to audio generation through text,'' \emph{Advances in Neural Information Processing Systems}, vol.~37, pp. 101\,337--101\,366, 2024.

\bibitem{Frieren}
Y.~Wang, W.~Guo, R.~Huang, J.~Huang, Z.~Wang, F.~You, R.~Li, and Z.~Zhao, ``Frieren: Efficient video-to-audio generation with rectified flow matching,'' \emph{arXiv e-prints}, pp. arXiv--2406, 2024.

\bibitem{foleycrafter}
Y.~Zhang, Y.~Gu, Y.~Zeng, Z.~Xing, Y.~Wang, Z.~Wu, and K.~Chen, ``Foleycrafter: Bring silent videos to life with lifelike and synchronized sounds,'' \emph{arXiv preprint arXiv:2407.01494}, 2024.

\bibitem{V2A-Mapper}
H.~Wang, J.~Ma, S.~Pascual, R.~Cartwright, and W.~Cai, ``V2a-mapper: A lightweight solution for vision-to-audio generation by connecting foundation models,'' in \emph{Proceedings of the AAAI Conference on Artificial Intelligence}, vol.~38, no.~14, 2024, pp. 15\,492--15\,501.

\bibitem{klingfoley}
J.~Wang, X.~Zeng, C.~Qiang, R.~Chen, S.~Wang, L.~Wang, W.~Zhou, P.~Cai, J.~Zhao, N.~Li \emph{et~al.}, ``Kling-foley: Multimodal diffusion transformer for high-quality video-to-audio generation,'' \emph{arXiv preprint arXiv:2506.19774}, 2025.

\bibitem{multifoley}
Z.~Chen, P.~Seetharaman, B.~Russell, O.~Nieto, D.~Bourgin, A.~Owens, and J.~Salamon, ``Video-guided foley sound generation with multimodal controls,'' in \emph{Proceedings of the Computer Vision and Pattern Recognition Conference}, 2025, pp. 18\,770--18\,781.

\bibitem{MMAudio}
H.~K. Cheng, M.~Ishii, A.~Hayakawa, T.~Shibuya, A.~Schwing, and Y.~Mitsufuji, ``Taming multimodal joint training for high-quality video-to-audio synthesis,'' \emph{arXiv e-prints}, pp. arXiv--2412, 2024.

\bibitem{sd3}
P.~Esser, S.~Kulal, A.~Blattmann, R.~Entezari, J.~M{\"u}ller, H.~Saini, Y.~Levi, D.~Lorenz, A.~Sauer, F.~Boesel \emph{et~al.}, ``Scaling rectified flow transformers for high-resolution image synthesis,'' in \emph{Forty-first international conference on machine learning}, 2024.

\bibitem{synchformer}
V.~Iashin, W.~Xie, E.~Rahtu, and A.~Zisserman, ``Synchformer: Efficient synchronization from sparse cues,'' in \emph{ICASSP 2024-2024 IEEE International Conference on Acoustics, Speech and Signal Processing (ICASSP)}.\hskip 1em plus 0.5em minus 0.4em\relax IEEE, 2024, pp. 5325--5329.

\bibitem{vggsound}
H.~Chen, W.~Xie, A.~Vedaldi, and A.~Zisserman, ``Vggsound: A large-scale audio-visual dataset,'' in \emph{ICASSP 2020-2020 IEEE International Conference on Acoustics, Speech and Signal Processing (ICASSP)}.\hskip 1em plus 0.5em minus 0.4em\relax IEEE, 2020, pp. 721--725.

\bibitem{audiocaps}
C.~D. Kim, B.~Kim, H.~Lee, and G.~Kim, ``Audiocaps: Generating captions for audios in the wild,'' in \emph{Proceedings of the 2019 Conference of the North American Chapter of the Association for Computational Linguistics: Human Language Technologies, Volume 1 (Long and Short Papers)}, 2019, pp. 119--132.

\bibitem{wavcaps}
X.~Mei, C.~Meng, H.~Liu, Q.~Kong, T.~Ko, C.~Zhao, M.~D. Plumbley, Y.~Zou, and W.~Wang, ``Wavcaps: A chatgpt-assisted weakly-labelled audio captioning dataset for audio-language multimodal research,'' \emph{IEEE/ACM Transactions on Audio, Speech, and Language Processing}, vol.~32, pp. 3339--3354, 2024.

\bibitem{GreatestHits}
A.~Owens, P.~Isola, J.~McDermott, A.~Torralba, E.~H. Adelson, and W.~T. Freeman, ``Visually indicated sounds,'' in \emph{Proceedings of the IEEE conference on computer vision and pattern recognition}, 2016, pp. 2405--2413.

\bibitem{flowmatching}
Y.~Lipman, R.~T. Chen, H.~Ben-Hamu, M.~Nickel, and M.~Le, ``Flow matching for generative modeling,'' \emph{arXiv preprint arXiv:2210.02747}, 2022.

\bibitem{adaln}
E.~Perez, F.~Strub, H.~De~Vries, V.~Dumoulin, and A.~Courville, ``Film: Visual reasoning with a general conditioning layer,'' in \emph{Proceedings of the AAAI conference on artificial intelligence}, vol.~32, no.~1, 2018.

\bibitem{audioldm2}
H.~Liu, Y.~Yuan, X.~Liu, X.~Mei, Q.~Kong, Q.~Tian, Y.~Wang, W.~Wang, Y.~Wang, and M.~D. Plumbley, ``Audioldm 2: Learning holistic audio generation with self-supervised pretraining,'' \emph{IEEE/ACM Transactions on Audio, Speech, and Language Processing}, vol.~32, pp. 2871--2883, 2024.

\bibitem{tango2}
N.~Majumder, C.-Y. Hung, D.~Ghosal, W.-N. Hsu, R.~Mihalcea, and S.~Poria, ``Tango 2: Aligning diffusion-based text-to-audio generations through direct preference optimization,'' in \emph{Proceedings of the 32nd ACM International Conference on Multimedia}, 2024, pp. 564--572.

\bibitem{makeanaudio2}
J.~Huang, Y.~Ren, R.~Huang, D.~Yang, Z.~Ye, C.~Zhang, J.~Liu, X.~Yin, Z.~Ma, and Z.~Zhao, ``Make-an-audio 2: Temporal-enhanced text-to-audio generation,'' \emph{arXiv preprint arXiv:2305.18474}, 2023.

\bibitem{GenAU-Large}
M.~Haji-Ali, W.~Menapace, A.~Siarohin, G.~Balakrishnan, and V.~Ordonez, ``Taming data and transformers for audio generation,'' \emph{arXiv preprint arXiv:2406.19388}, 2024.

\bibitem{CondFoleyGen}
Y.~Du, Z.~Chen, J.~Salamon, B.~Russell, and A.~Owens, ``Conditional generation of audio from video via foley analogies,'' in \emph{Proceedings of the IEEE/CVF Conference on Computer Vision and Pattern Recognition}, 2023, pp. 2426--2436.

\bibitem{sketch2sound}
H.~F. Garc{\'\i}a, O.~Nieto, J.~Salamon, B.~Pardo, and P.~Seetharaman, ``Sketch2sound: Controllable audio generation via time-varying signals and sonic imitations,'' in \emph{ICASSP 2025-2025 IEEE International Conference on Acoustics, Speech and Signal Processing (ICASSP)}.\hskip 1em plus 0.5em minus 0.4em\relax IEEE, 2025, pp. 1--5.

\bibitem{dacvae}
R.~Kumar, P.~Seetharaman, A.~Luebs, I.~Kumar, and K.~Kumar, ``High-fidelity audio compression with improved rvqgan,'' \emph{Advances in Neural Information Processing Systems}, vol.~36, pp. 27\,980--27\,993, 2023.

\bibitem{voicebox}
M.~Le, A.~Vyas, B.~Shi, B.~Karrer, L.~Sari, R.~Moritz, M.~Williamson, V.~Manohar, Y.~Adi, J.~Mahadeokar \emph{et~al.}, ``Voicebox: Text-guided multilingual universal speech generation at scale,'' \emph{Advances in neural information processing systems}, vol.~36, pp. 14\,005--14\,034, 2023.

\bibitem{clip}
A.~Radford, J.~W. Kim, C.~Hallacy, A.~Ramesh, G.~Goh, S.~Agarwal, G.~Sastry, A.~Askell, P.~Mishkin, J.~Clark \emph{et~al.}, ``Learning transferable visual models from natural language supervision,'' in \emph{International conference on machine learning}.\hskip 1em plus 0.5em minus 0.4em\relax PmLR, 2021, pp. 8748--8763.

\bibitem{latent-inversion}
G.~Le~Lan, B.~Shi, Z.~Ni, S.~Srinivasan, A.~Kumar, B.~Ellis, D.~Kant, V.~Nagaraja, E.~Chang, W.-N. Hsu \emph{et~al.}, ``High fidelity text-guided music generation and editing via single-stage flow matching,'' \emph{arXiv e-prints}, pp. arXiv--2407, 2024.

\bibitem{vta-ldm}
M.~Xu, C.~Li, X.~Tu, Y.~Ren, R.~Chen, Y.~Gu, W.~Liang, and D.~Yu, ``Video-to-audio generation with hidden alignment,'' \emph{arXiv preprint arXiv:2407.07464}, 2024.

\bibitem{cfg}
J.~Ho and T.~Salimans, ``Classifier-free diffusion guidance,'' \emph{arXiv preprint arXiv:2207.12598}, 2022.

\bibitem{vggish}
S.~Hershey, S.~Chaudhuri, D.~P. Ellis, J.~F. Gemmeke, A.~Jansen, R.~C. Moore, M.~Plakal, D.~Platt, R.~A. Saurous, B.~Seybold \emph{et~al.}, ``Cnn architectures for large-scale audio classification,'' in \emph{2017 ieee international conference on acoustics, speech and signal processing (icassp)}.\hskip 1em plus 0.5em minus 0.4em\relax IEEE, 2017, pp. 131--135.

\bibitem{inceptionscore}
T.~Salimans, I.~Goodfellow, W.~Zaremba, V.~Cheung, A.~Radford, and X.~Chen, ``Improved techniques for training gans,'' \emph{Advances in neural information processing systems}, vol.~29, 2016.

\bibitem{panns}
Q.~Kong, Y.~Cao, T.~Iqbal, Y.~Wang, W.~Wang, and M.~D. Plumbley, ``Panns: Large-scale pretrained audio neural networks for audio pattern recognition,'' \emph{IEEE/ACM Transactions on Audio, Speech, and Language Processing}, vol.~28, pp. 2880--2894, 2020.

\bibitem{clap}
Y.~Wu, K.~Chen, T.~Zhang, Y.~Hui, T.~Berg-Kirkpatrick, and S.~Dubnov, ``Large-scale contrastive language-audio pretraining with feature fusion and keyword-to-caption augmentation,'' in \emph{ICASSP 2023-2023 IEEE International Conference on Acoustics, Speech and Signal Processing (ICASSP)}.\hskip 1em plus 0.5em minus 0.4em\relax IEEE, 2023, pp. 1--5.

\bibitem{imagebind}
R.~Girdhar, A.~El-Nouby, Z.~Liu, M.~Singh, K.~V. Alwala, A.~Joulin, and I.~Misra, ``Imagebind: One embedding space to bind them all,'' in \emph{Proceedings of the IEEE/CVF conference on computer vision and pattern recognition}, 2023, pp. 15\,180--15\,190.

\bibitem{vggsound-sync}
H.~Chen, W.~Xie, T.~Afouras, A.~Nagrani, A.~Vedaldi, and A.~Zisserman, ``Audio-visual synchronisation in the wild,'' \emph{arXiv preprint arXiv:2112.04432}, 2021.

\bibitem{passt}
K.~Koutini, J.~Schl{\"u}ter, H.~Eghbal-Zadeh, and G.~Widmer, ``Efficient training of audio transformers with patchout,'' \emph{arXiv preprint arXiv:2110.05069}, 2021.

\end{thebibliography}
\end{document}